\documentclass[a4paper]{article}

\usepackage{INTERSPEECH2021}
\usepackage{xcolor}
\usepackage{multirow}
\usepackage{hyperref}

\title{Many-Speakers Single Channel Speech Separation\\ with Optimal Permutation Training}
\name{Shaked Dovrat$^1$*, Eliya Nachmani$^{1,2}$*, Lior Wolf$^{1}$}

\address{
  $^1$Blavatnik School of Computer Science, Tel Aviv University, Israel\\
  $^2$Facebook AI Research, Israel}
\email{shaked.dovrat@gmail.com, enk100@gmail.com, wolf@cs.tau.ac.il}

\begin{document}
\maketitle
{\let\thefootnote\relax\footnote{{*Equal contribution}}}
\vspace{-.55cm}
\begin{abstract}
    Single channel speech separation has experienced great progress in the last few years. However, training neural speech separation for a large number of speakers (e.g., more than 10 speakers) is out of reach for the current methods, which rely on the Permutation Invariant Training (PIT).
    In this work, we present a permutation invariant training that employs the Hungarian algorithm in order to train with an $O(C^3)$ time complexity, where $C$ is the number of speakers, in comparison to $O(C!)$ of PIT based methods. Furthermore, we present a modified architecture that can handle the increased number of speakers. Our approach separates up to $20$ speakers and improves the previous results for large $C$ by a wide margin.
\end{abstract}

\section{Introduction}
Single channel speech separation is a fundamental problem in the speech process, which has seen tremendous advances in the last years. The main neural architectures can be divided into two categories: (i) Spectral based~\cite{wang2019deep,li2019spectral, wang2018end} and (ii) time domain based~\cite{luo2018tasnet, nachmani2020voice, zeghidour2020wavesplit}. Currently, the latter category leads with respect to the obtained accuracy.

Since the order of the speakers at the output of the neural network is arbitrary, a permutation invariant training is performed. Most of the neural architecture for speech separation use the permutation invariant training (PIT) loss \cite{yu2017permutation} or its extension to the utterance level (uPIT)~\cite{kolbaek2017multitalker}. Both variants have a computational complexity of $O(C!)$, where $C$ is the number of the speakers. As a result, it is not feasible to run PIT on more than ten speakers. 

In this work, following \cite{kanda2020serialized} and \cite{ma2020monaural}, we propose a novel method to train a large number of speakers with a lower complexity of $O(C^3)$, by using the Hungarian algorithm. The Hungarian algorithm is able to find the optimal permutation in terms of the minimal sum of pairwise losses, thus matching between pairs of output- and target-signals. In order to enable the separation network to deal with a large number of speakers, we further introduce an architecture that combines two distinct approaches to separation networks, LSTM and dilated convolutional layers.

In our experiments, our method separates up to $20$ speakers, which, as far as we can ascertain, is twice the number tackled by any existing method. Moreover, we show that our method improves the previous state of the art separation results for separating $5$ and $10$ speakers. 

%\vspace{-0.25cm}
\section{Related Work}

Signal channel speech separation was explored using classical approaches~\cite{martin2018single,ernst2018speech} and, more recently, using deep learning methods. In \cite{erdogan2015phase} an LSTM neural network with a phase sensitive loss function was introduced. An improvement in SDR was demonstrated on the CHiME-2 \cite{vincent2013second} dataset. In \cite{hershey2016deep} a neural separation network with a clustering-based embedding was introduced, presenting results for the separation of two speakers and introducing the WSJ-2mix dataset that was extensively used by followup work. This work was further extended  in \cite{isik2016single} by  extracting an embedding of spectrogram segments and estimating a mask for the separation part. Results were provided for two and three speakers and an SDR improvement of $10.3$ dB and $7.1$ dB for WSJ-2mix and WSJ-3mix was obtained. In \cite{chen2017deep}, the neural separator network was introduced. Attractor points in the embedding space were used to obtain the time-frequency bins for each speaker. The improvement on the WSJ-mix dataset was by 5.49\%. 

Luo et al. \cite{luo2018tasnet} introduced TasNet, which is a time domain encoder-decoder neural architecture for the single channel speech separation problem. They show a results of 11.1 SDR improvement for WSJ-2mix dataset over the state of the art. Wang et al. \cite{wang2018end} proposed a neural architecture that separates the speakers in both the time and the frequency domains simultaneously. They presented an improvement of 13.2 SDR on the WSJ-2mix dataset. The work of \cite{luo2018tasnet} further improved the architecture and introduced ConvTasNet \cite{luo2019conv}, which employed a dilated convolutional neural network, showing an SDR improvement 15.6 dB for the WSJ-2mix dataset. 

Another improvement with LSTM network was introduced in \cite{luo2019dual}, where the dual-path recurrent neural network (DPRNN) architecture was employed to model extremely long sequences. They showed SDR improvement of 18.08 dB on WSJ-2mix dataset. In \cite{nachmani2020voice} a separation network with $MulCat$ blocks was introduced. The proposed method also removed the masking sub-network, leading to an improvement of 20.12 dB to WSJ-2mix dataset. Furthermore, the WSJ-mix dataset was extended to include mixtures of $5$ speakers, where the SDR improvement was 10.6 dB. \cite{shi2020toward} combined DPRNN and TasNet and for the WSJ-5mix dataset they showed an SDR improvement of 10.41dB, and 11.14dB SDR improvement for online remixing. Zeghidour et al. \cite{zeghidour2020wavesplit} introduced a neural separation network that infers a representation to each speaker, by performing clustering, and used it to separate the mixture. They show an SDR improvement of 22.2 dB for WSJ-2mix dataset. 

Since our work builds upon the $MulCat$ network architecture~\cite{nachmani2020voice}, we will recap its major components. The network consists of an encoder, a separation module and a decoder. The encoder and decoder are simple 1D convolutions. The separation module starts with a chunking module which cuts the signal into chunks in time. Then, a series of doubled $MulCat$ blocks is applied. During training, a multi-scale loss is employed- after each doubled $MulCat$ block the activations are reconstructed by the decoder into audio signals and fed into the loss function. The method also uses the Scale-Invariant Signal-to-Noise Ratio (SI-SNR) loss, which is a slight improvement to the traditional SDR loss.

Another line of work for speech separation uses beamformers and introduces an extension of the minimum variance distortionless response (MVDR)~\cite{markovich2009multichannel}. A neural beamformer was introduced in \cite{sainath2015speaker} and further improved in \cite{luo2019fasnet} for the speech separation problem. A follow up work introduced the linearly constrained minimum variance (LCMV) beamformer \cite{laufer2020global}. 

In \cite{tachibana2020towards} a SinkPIT loss is introduced. They proposed a variant of the PIT loss, which is based on Sinkhorn’s matrix balancing algorithm. They reduce the complexity of the PIT loss from $O(C!)$ to $O(kC^2)$, where $k$ is set to $200$. It is important to note that the chosen permutation is only an approximation of the optimal permutation. In another work, a probabilistic-PIT loss which considers the output permutation as discrete latent random variable was introduced~\cite{yousefi2019probabilistic}. \cite{kanda2020serialized} and \cite{ma2020monaural} noted that it is possible to use the Hungarian algorithm to find the best permutation for source separation.

\vspace{-0.1cm}
\subsection{Hungarian Algorithm}
The \textit{linear sum assignment} problem (also known as the \textit{assignment problem}) is the task of assigning $C$ agents to do $C$ tasks, such that each agent is assigned to exactly one task, and the total cost for the agents performing the tasks is minimal. In other words, given a $C$-by-$C$ matrix of costs, $M$, one for each agent-task pair, find a permutation $\pi$ of the agents, such that the sum of costs of paired agents and tasks is minimal:

\vspace{-3mm}
\begin{equation}
\pi = \underset{\pi \in \Pi_C}{\operatorname{argmin}}\sum_{i=1}^n M_{i,\pi(i)}
\end{equation}
\vspace{-1mm}

A naive solution is to iterate over all $C!$ possible permutations. Fortunately, an optimal and polynomial-time algorithm that solves the assignment problem was proposed in $1955$ by Harold Kuhn \cite{kuhn1955hungarian}, reviewed in $1957$ by James Munkres \cite{munkres1957algorithms} and is mostly known by the name the \textit{Hungarian Algorithm}. The initial time complexity of the algorithm was $O(C^4)$ and it was modified later on to a time complexity of $O(C^3)$~\cite{tomizawa1971some}.

Simply put, the algorithm starts by trying to find an obvious permutation. If that fails, it goes on to make modifications to the input matrix, in order to find a valid and optimal permutation. The number of the modification iterations needed to find the solution is indicative of how well the permutation fits the data compared to alternative permutations. In other words, the more iterations needed until convergence, the less significant is the optimal permutation. 

%\vspace{-0.25cm}
\section{Method}

\begin{figure*}[t]
\centering
\begin{tabular}{c@{~}c}

\includegraphics[width=.6\textwidth,height=.3\textheight,keepaspectratio]{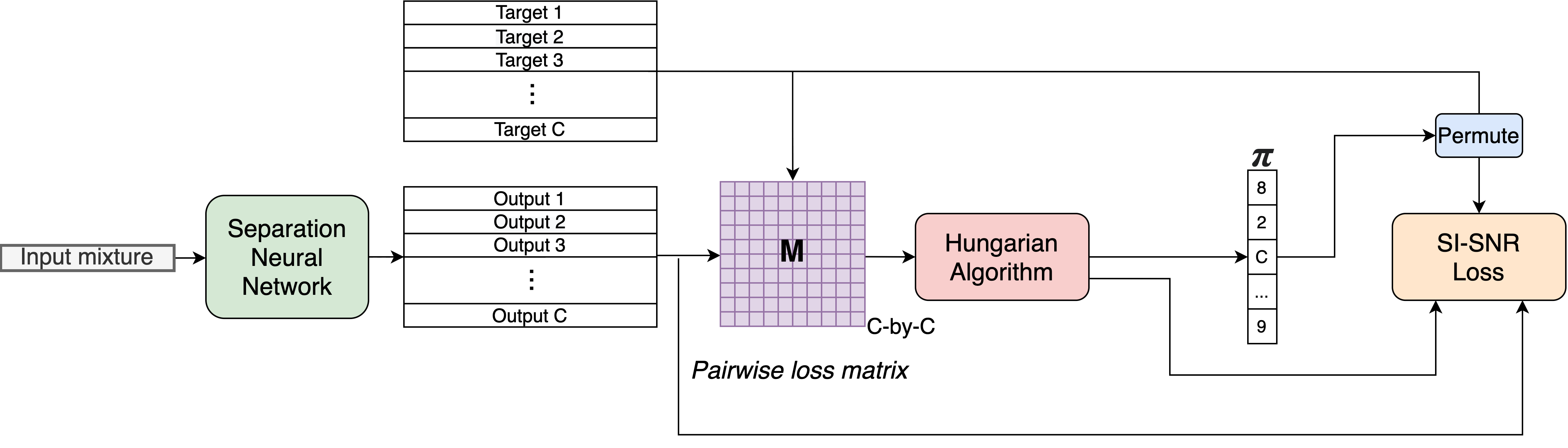}\\
(i) \\
\includegraphics[width=.6\textwidth,height=.3\textheight,keepaspectratio]{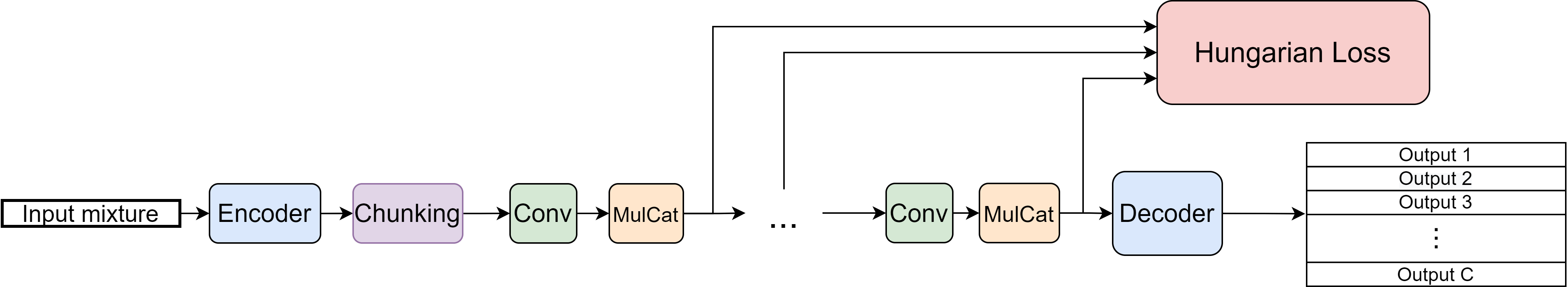}\\
(ii) \\
\end{tabular}
\smallskip
\caption{(i) The proposed Hungarian Loss. $M$, a $C$-by-$C$ matrix of SI-SNR losses between output and target pairs is computed. $M$ is fed into the Hungarian algorithm, which efficiently finds the optimal permutation of target signals, $\pi$. (ii) The proposed separation network architecture. The novel components are the added Conv blocks and the Hungarian loss, which replaces the PIT loss.}
\vspace{-.41cm}
\label{fig:overall}
\end{figure*}

This work extends the work in \cite{nachmani2020voice}, through a number of new contributions. First, we introduce the Hungarian Loss which replaces the PIT loss and gives an optimal solution to the permutation issue with a much lower time complexity, $O(C^3)$, which allows to train separation networks for many speakers. Second, we introduce a new network architecture that uses stacked dilated convolutions before each pair of $MulCat$ blocks of \cite{nachmani2020voice}. The overall method, is depicted in Figure \ref{fig:overall}.

\vspace{-0.1cm}
\subsection{Hungarian Loss}
A single-channel speech separation network takes an audio signal that contains a mixture of $C$ speakers speaking and outputs $C$ audio signals, each optimized to contain a separate speaker.

During training, the network outputs the separated audio signals in an arbitrary order. Thus, in order to compute a meaningful loss, an alignment, i.e. a permutation, needs to be recovered between the outputs of the network and the separated target signals.  One way to find the right permutation is to iterate over all possible $C!$ permutations and choose the one which gives the lowest mean loss value on the pairs (i.e. PIT). The computational cost of PIT is unnoticeable when $C$ is small, in comparison to the other parts of the network. However, it makes training on a large number of speakers impossible. For instance, for 20 speakers PIT needs to check $20!\approx2.4\times 10^{18}$ different permutations).

To address this issue, we formulate the task of finding the permutation which minimizes the loss function as a \textit{linear sum assignment} problem. Given the $C$ output signals and $C$ target signals, we calculate the pairwise loss value, $\hat{\ell}(s_i, \hat{s}_{j})$, on every pair of output ($\hat{s}_{j}$) and target ($s_i$) signals, which gives an $C$-by-$C$ matrix of losses, $M$. Next, we assign each output with a unique target and vice-versa. Such assignment is equivalent to choosing $C$ elements of the matrix, such that each chosen element is in a different row and column from all others. An optimal assignment minimizes the sum of values of the chosen elements. The PIT loss can then be viewed as a brute-force solution to this problem, iterating over all possible solutions:

\vspace{-0.3cm}
\begin{equation}
\ell(s, \hat{s}) = \min_{\pi \in \Pi_C}~\frac{1}{C}\sum_{i=1}^C \hat{\ell}(s_i, \hat{s}_{\pi(i)})
\end{equation}
\vspace{-0.1cm}

By running the Hungarian Algorithm on $M$, we efficiently find the optimal permutation in polynomial time instead of the brute-force, factorial-time PIT. Note that the assignment algorithm does not need to be differential, since we find a permutation of the targets by which we calculate the loss, meaning that the process is separate from the backwards calculation of gradients.

\vspace{-0.1cm}
\subsection{Model}
We shift our focus to solving the task of separating mixtures of many ($C\geq10$) speakers. As $C$ increases, the task of separating the mixtures becomes more challenging. Thus, we propose a new and suitable network architecture.

For this end, we modify the $MulCat$-based architecture~\cite{nachmani2020voice} by adding stacked dilated convolutions before each pair of $MulCat$ blocks. In addition, we increased some of the network's hyper-parameters to achieve a larger capacity needed for the harder tasks. The dilated convolutions scheme is borrowed from \cite{luo2019conv}: We use $8$ \textit{1-D Conv block}s stacked on top of each other, with dilation factors $2^{i-1}$ for $i\in\{1,2,...,8\}$. This corresponds to a single column in the separation module of \cite{luo2019conv}.

The rest of the model is in accordance with \cite{nachmani2020voice}, i.e. using the same encoder, chunking, $MulCat$ blocks, decoder and SI-SNR loss. We also adopt the multi-scale loss scheme, which applies the loss function after each double MulCat block, instead of just the last. We tuned some hyper-parameters of the architecture to accompany the harder tasks when $C$ is large: $N$, number of features, was increased from $128$ to $256$. $L$, the encoder's kernel size was increased from $8$ to $16$. $H$, the number of hidden units in the LSTMs was increased from $128$ to $256$  and finally $R$, the number of double $MulCat$ blocks was increased from $6$ to $7$. A similar hyper-parameters adjustment was done in \cite{tachibana2020towards}.

\begin{figure}[h]
    \centering

    \includegraphics[width=.3\textwidth,height=.3\textheight,keepaspectratio]{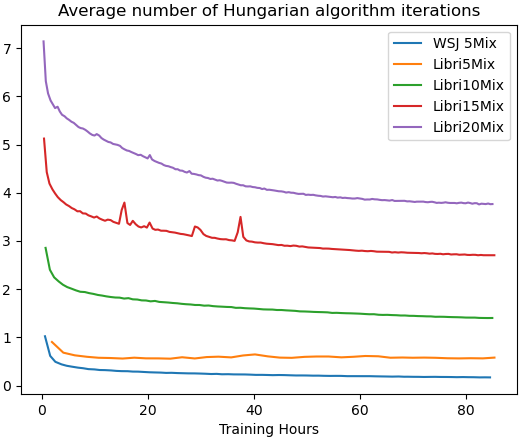} 
    \caption{The average number of iterations per example performed by the Hungarian algorithm. The average is decreasing as training progresses, which indicates that the separation is improving. This also means that the runtime of the Hungarian algorithm is getting even shorter.}
    \label{fig:HungaryWeight}
    \vspace{-.41cm}
\end{figure}
\vspace{-2mm}

\begin{figure*}[t]
%\vspace{-1.5cm}
\centering
\begin{tabular}{c}
\includegraphics[width=0.9\linewidth]{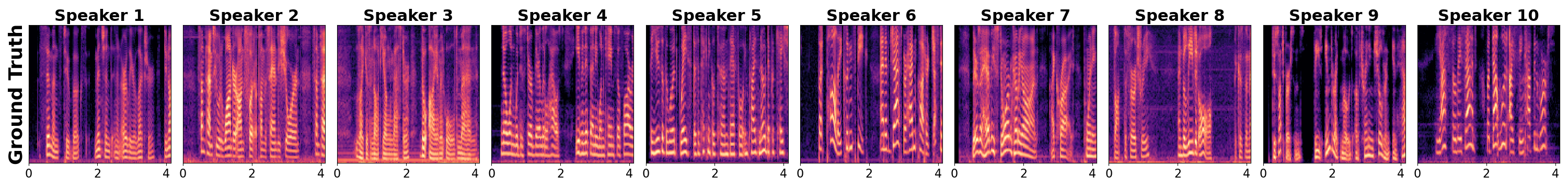} \\
\includegraphics[width=0.9\linewidth]{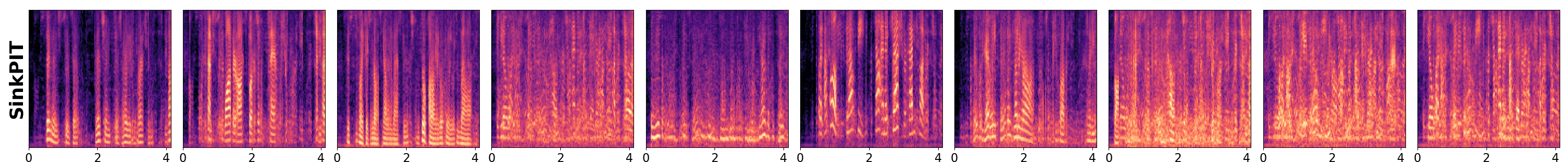} \\
\includegraphics[width=0.9\linewidth]{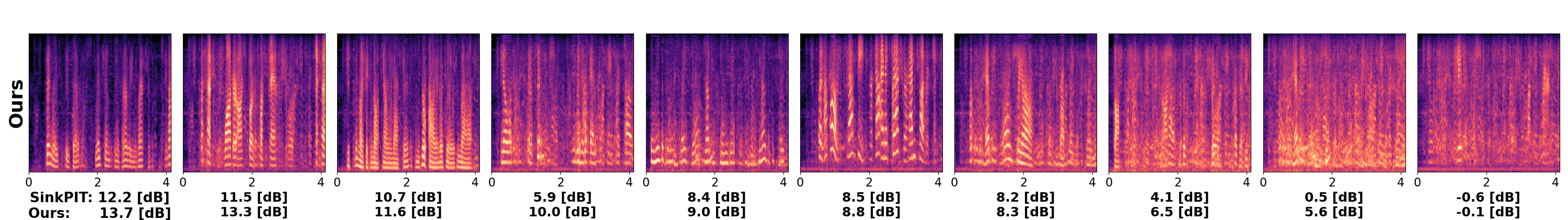} \\
\end{tabular}
\caption{Separation results for a mixture of $10$ speakers. First row: mel spectrogram of ground truth signals. Second row: mel spectrogram of SinkPIT signals. Third row: mel spectrogram of the outputs of our method. The last row shows the SI-SDR improvement for the SinkPIT method and our method. The x-axis is sorted by the SI-SDRi in descending order.}
\label{fig:spec}
\vspace{-.41cm}
\end{figure*}

\begin{figure}[h]
%\centering
\hspace*{-0.6cm}
\begin{tabular}{cc}
\includegraphics[width=0.5\linewidth]{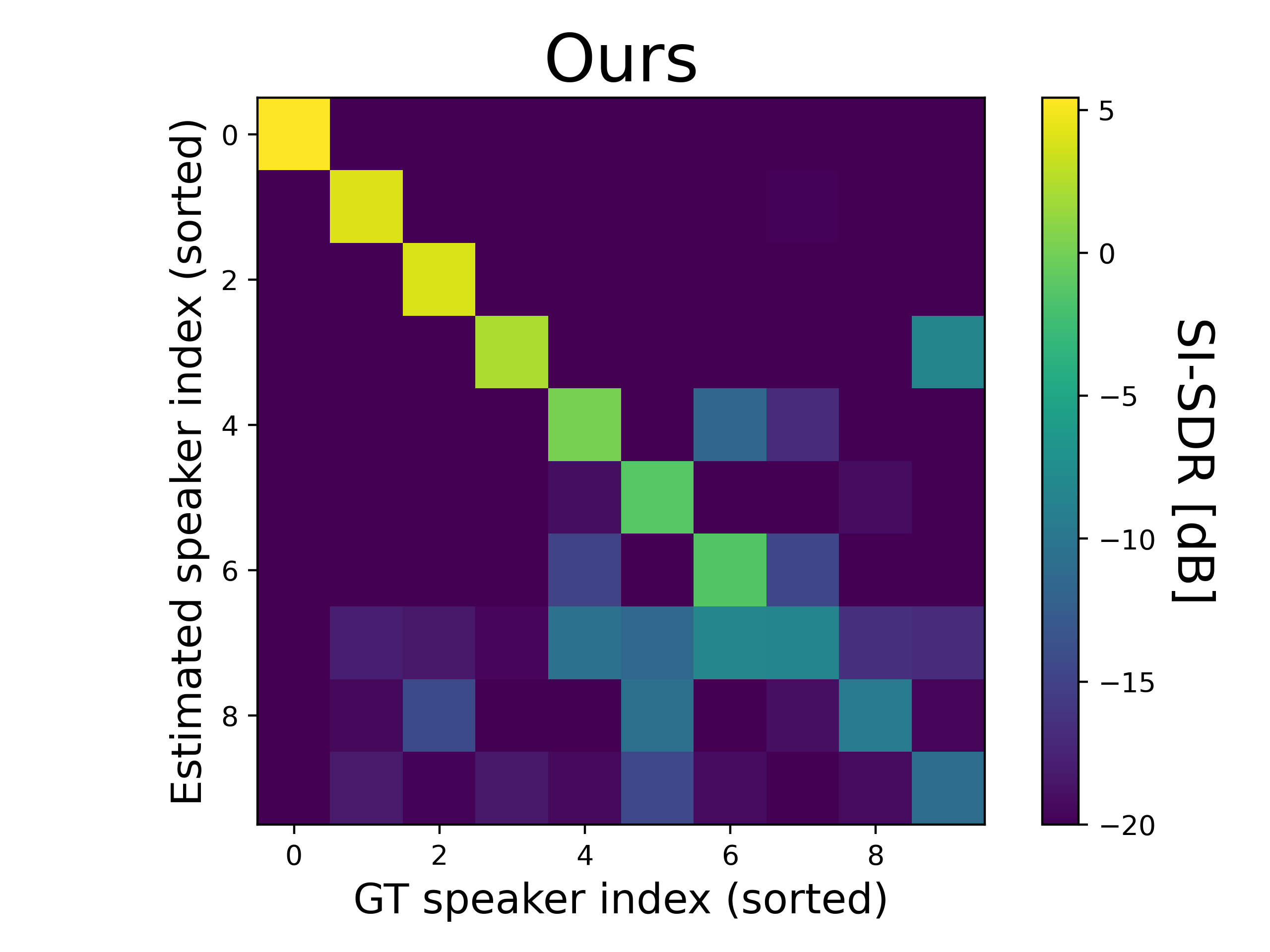} &
\includegraphics[width=0.5\linewidth]{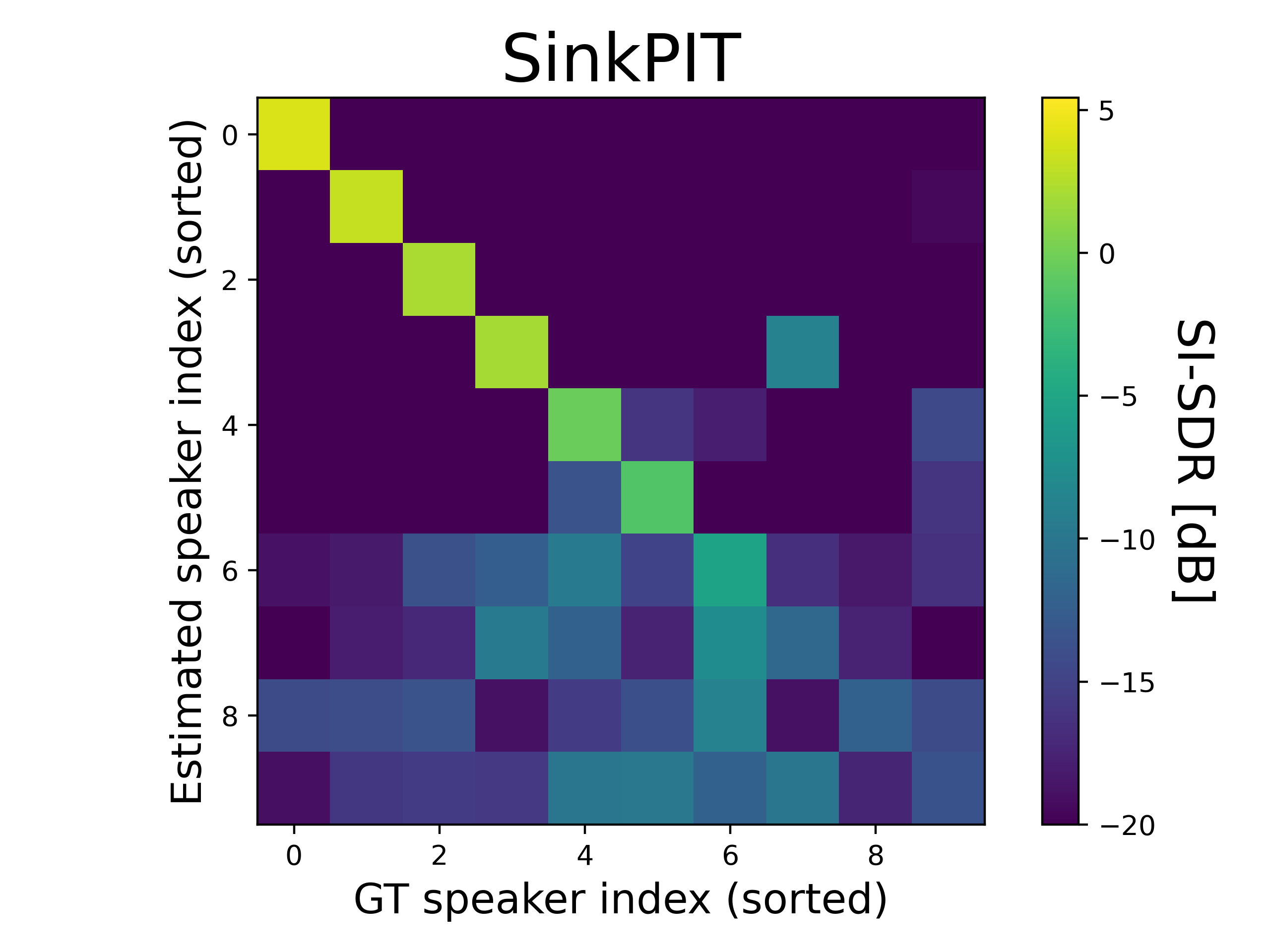} \\
(a) & (b) \\
\end{tabular}
\vspace{-.25cm}
\caption{The pairwise SI-SDR matrix $M$, sorted in descending order. For the same input sample as in Figure \ref{fig:spec}. (a) Our method. (b) SinkPIT \cite{tachibana2020towards}.}
\vspace{-.65cm}
\label{fig:mat}
\end{figure}

%\vspace{-0.15cm}
\section{Experiments}
\noindent{\bf Comparison to state of the art\quad}
We show results on datasets derived from WSJ corpus \cite{garofolo1993csr} and LibriSpeech \cite{panayotov2015LibriSpeech}. For WSJ, we use the 5-speaker mix, introduced in \cite{nachmani2020voice}, which uses the same procedure as in \cite{hershey2016deep}, i.e. $30$ hours of speech from the training set si$\_$tr$\_$s were used to create the training and validation sets. The five speakers were randomly chosen and combined with random SNR values between $0-5$ dB. The test set is created from si$\_$et$\_$s and si$\_$dt$\_$s with 16 speakers, that differ from the speakers of the training set. For LibriSpeech, we use the LibriMix \cite{cosentino2020librimix} datasets with mixes of 5, 10, 15 and 20 speakers. LibriMix offers mixtures of 2 and 3 speakers from LibriSpeech, and we used the given scripts to create mixtures of 5, 10, 15 and 20 speakers. Some modifications to the script needed to be done. These can be found online*. We used LibriMix's given parameters to get a sample rate of $8$Khz, clean mixtures (no noise added) and each sample is cut in length, according to the minimal length sample in the mixture. We also use the augmentation process as in \cite{tachibana2020towards}.

{\let\thefootnote\relax\footnote{{*\url{https://github.com/ShakedDovrat/LibriMix}}}}

A separate model is trained for each dataset, with the corresponding number of output channels. Training was done using the Adam optimizer \cite{kingma2014adam}, with batch size $32$ and a learning rate of $1e-3$ which was multiplied by $0.95$ every two epochs. During training, each sample is cut into 4-second segments. Table~\ref{tab:results} compares the results of our model to other models, using the SI-SDRi metrics.  As can be seen, our model outperforms the other methods by a large margin. For Libri-5mix and Libri-10mix, we improve the previous results by $1.89$dB and $1.33$dB respectively. Interestingly, the previous state of the art results for $Libri5Mix$ is obtained with $MulCut$ \cite{nachmani2020voice}, whereas for $Libri10Mix$ it is obtained by \cite{tachibana2020towards}, which uses SinkPIT. This is due to the fact that the PIT loss in $MulCut$ for 10 speakers is prohibitive. For WSJ-5mix, the SDR of our method improves by more than $2$dB over the previous method. We are the first to present results for the $Libri15Mix$ and $Libri20Mix$ datasets and running previous work on it is not practical. Sample results are shared online~\url{https://shakeddovrat.github.io/hungarian/}.

\noindent{\bf Hungarian loss vs. alternative losses\quad} In order to show the benefits of using the Hungarian algorithm as opposed to PIT, we show the training duration of an epoch of each dataset, depicted in Table \ref{tab:timing}. As shown, on a $5$ speaker mix both methods take about the same time. However, for $C=10$, the Hungarian method is about $9$ times faster on our model. For $C\geq15$, the Hungarian method is still fast, while PIT is unusable as it failed to complete a single epoch after days of running. To our knowledge, the SinkPIT can run on $20$ speakers but only find an approximation to the optimal permutation, whereas our method is both optimal and fast (the source code was not published). 

In Figure~\ref{fig:HungaryWeight}, we present the average number of iterations per example that were performed by the Hungarian algorithm for the various datasets. We can observe two phenomena: (i) As the training proceeds and the neural network improves the separation performance, the average number of iterations is decreased. (ii) The average number of iterations is lower when the number of speakers in the mixture is lower. Both of these observations came from the fact that the Hungarian algorithm needs no iterations to converge when the outputs or the network are mixed or noisy.

In Figure~\ref{fig:spec}, we plot the mel spectrogram for typical samples as in \cite{tachibana2020towards}. This is shown for a mixture of $10$ speakers, from the $Libri10Mix$ dataset. As can be seen in speakers $4$ and $8$ our method provide a cleaner mel spectrogram, Furthermore, the SI-SDRi difference is 4.9 dB and 2.4 dB for speakers $4$ and $8$ respectively (there is a significant improvement in all speakers). 

In Figure \ref{fig:mat}, we plot the pairwise SI-SDR negative matrix $M$ sorted in descending order, in comparison to the results of the SinkPIT system~\cite{tachibana2020towards}. Evidently, the entropy in our method is lower, especially in the last rows, i.e., it has much less confusion compared to the baseline method.

\noindent{\bf Ablation study\quad} We run ablation analysis in order to understand the contribution of each component of our method. The results are summarized in Table \ref{tab:ablation}, where -C means without the 1D convolutions and -H means using the PIT loss instead of the Hungarian loss. As can be seen, adding the dilated convolution to the architecture improves the performance. Moreover, without the Hungarian algorithm, it is practically impossible to train on $15$ or more speakers. For $10$ speakers, using PIT drastically diminishes performance due to longer training time. On the other hand, for $5$ speakers, the PIT and Hungarian have similar performance since they both found the optimal permutation in a similar runtime. 

%\vspace{-0.25cm}
\section{Conclusions}
In this work, we provide a method for single channel sound separation for a large number of sources. Our method is the first work to show that one can separate a mixture of $20$ speakers from a single channel recording. Our solution is based on the Hungarian algorithm, which efficiently finds the optimal permutation from the $C!$ possible permutations and on a new network architecture that adds stacked 1-D convolutions and added capacity to the state of the art architecture.

\begin{table}[]
\caption{SDR improvement performance of various models versus number of speakers and datasets. 'X' results indicated simulation that failed to complete a single epoch, due to high complexity PIT loss. \textit{x}Mix is LibriMix with \textit{x} speakers.}
\label{tab:results}
\begin{tabular}{lc@{\hspace{1\tabcolsep}}c@{\hspace{1\tabcolsep}}c@{\hspace{1\tabcolsep}}c|c}
\toprule
Model       & \textbf{5Mix} & \textbf{10Mix} & \textbf{15Mix} & \textbf{20Mix} & \textbf{WSJ-5} \\
\midrule
ConvTasNet   & -                  & -                   & -                   & -                   & 6.8               \\
DPRNN   \cite{luo2019dual}  & -                  & -                   & -                   & -                   & 8.6    \\
MulCat  \cite{nachmani2020voice}                    & 10.83              & 4.74                & X                & X                & 10.6 \\
TasTas   \cite{shi2020toward}   & -              & -                & -                & -                & 11.14              \\
SinkPIT \cite{tachibana2020towards} & 9.39               & 6.45                & -                   & -                   & -      \\
\textbf{Ours}  & \textbf{12.72}     & \textbf{7.78}       & \textbf{5.66}       & \textbf{4.26}       & \textbf{13.22}  \\
\bottomrule
\end{tabular}
\end{table}

\begin{table}[]
\caption{Training run times of PIT and Hungarian algorithm in minutes per epoch. 'X' indicates simulation that failed to complete a single epoch, due to long run times of the PIT loss.}
\label{tab:timing}
\begin{tabular}{lccccc}
\toprule
\textbf{Dataset}                 & \textbf{\#Spkrs} & \textbf{\#Perms} & \textbf{PIT} & \textbf{Hungarian} \\
\midrule
WSJ-5mix                         & 5                   & 120                     & 65     & 62        \\
Libri-5Mix                       & 5                   & 120                     & 139    & 140       \\
Libri-10Mix                      & 10                  & $\approx3.6e6$          & 462    & 52        \\
Libri-15Mix                      & 15                  & $\approx1.3e12$         & X   & 36        \\
Libri-20Mix                      & 20                  & $\approx2.4e18$         & X   & 29        \\
\bottomrule
\end{tabular}
\end{table}

\begin{table}[]
\caption{Ablation analysis - SDR Performance for LibriMix and WSJ-mix datasets. \textit{x}Mix is LibriMix with \textit{x} speakers. \textit{"-C"} means without the 1D convolutions and \textit{"-H"} means using PIT instead of the Hungarian loss. 'X' indicates simulation that failed to complete a single epoch, due to long run times of the PIT loss.}
\label{tab:ablation}
\begin{tabular}{lcccc|c}
\toprule
\textbf{Model}                                    & \textbf{5Mix} & \textbf{10Mix} & \textbf{15Mix} & \textbf{20Mix} & \textbf{WSJ-5} \\
\midrule
Ours-C-H & 11.10               & 4.47                & X                & X     & 12.18 \\
Ours-H       & 12.53               & 4.84                & X                & X    & 13.07  \\
Ours-C           & 11.19              & 5.89                & 5.14                & 4.10     & 12.10    \\
\textbf{Ours}        & \textbf{12.72}     & \textbf{7.78}       & \textbf{5.66}       & \textbf{4.26}       & \textbf{13.22}  \\  
\bottomrule
\end{tabular}
\end{table}

%\vspace{-0.25cm}
\section{Acknowledgments}
 This project has received funding from the European Research Council (ERC) under the European Unions Horizon 2020 research and innovation programme (grant ERC CoG 725974). We thank Hideyuki Tachibana for the helpful discussion. The contribution of Eliya Nachmani is part of a Ph.D. thesis research conducted at Tel Aviv University.

\bibliographystyle{IEEEtran}
\bibliography{mybib}
\end{document}